\documentstyle[11pt,aaspp4]{article}

\slugcomment{To appear in the Astronomical Journal}

\lefthead{Liu, Petry, Impey \& Foltz}
\righthead{Quasars as Absorption Probes}

\begin{document}

\title{QUASARS AS ABSORPTION PROBES OF THE HUBBLE DEEP FIELD\footnote{
Observations reported here were obtained at the Multiple Mirror Telescope
Observatory, a facility operated jointly by the University of Arizona
and the Smithsonian Institution; and at
Kitt Peak National Observatory, National Optical 
Astronomy Observatories, operated by AURA Inc., under contract with
the National Science Foundation.}}

\author{Charles T. Liu}
\affil{Department of Astronomy, Columbia University, New York, NY 10027; and}
\affil{Department of Astrophysics, American Museum of Natural 
       History, New York, NY 10024}
\affil{email: cliu@astro.columbia.edu; cliu@amnh.org}

\author{Cathy E. Petry and Chris D. Impey}
\affil{Steward Observatory, The University of Arizona, Tucson, AZ 85721}
\affil{email: cpetry@as.arizona.edu, cimpey@as.arizona.edu}
\and
\author{Craig B. Foltz}
\affil{Multiple Mirror Telescope Observatory, The University of Arizona,
Tucson, AZ 85721}
\affil{email: cfoltz@as.arizona.edu}

\begin{abstract}

We present a catalog of 30 QSOs and their spectra, in the square
degree of sky centered on the northern Hubble Deep Field. These 
QSOs were selected by multicolor photometry and subsequently 
confirmed with spectroscopy. They range in magnitude from $17.6 
< B < 21.0$ and in redshift from $0.44 < z < 2.98$.  We also
include in the catalog an AGN with redshift z=0.135. Together, these
objects comprise a new grid of absorption probes which can be used to 
study the correlation between luminous galaxies, non-luminous 
halos and Lyman-$\alpha$ absorbers along the line of sight 
toward the Hubble Deep Field.

\end{abstract}


\section{INTRODUCTION} 

The Hubble Deep Field (HDF), with its unprecedented depth and its 
rich resource of complementary data, has opened new avenues for 
studying galaxy evolution and cosmology (Williams et al. 1996; Livio, 
Fall \& Madau 1998). The northern HDF no longer stands alone as the
subject of the deepest image of the sky ever made; it was recently
matched by deep Hubble Space Telescope (HST) observations of a
southern field (Williams et al. 1998). In this paper, we use the
term HDF to refer only to the northern field. Yet, no matter how 
deep any imaging survey might be, it can only reveal the luminous 
parts of galaxies, which comprise only 2-3\% of the material in the 
universe. An examination of the cold, diffuse and dark components of 
the universe along the line of sight toward the HDF would provide an 
important complement to the study of the luminous matter content
between $0 < z < 4$.

There are a number of benefits to an absorption survey, using distant 
QSOs as background probes. Material can be detected in absorption 
that would be impossible to detect in emission. For example, galaxy 
halos can be detected using the C\thinspace IV $\lambda\lambda$1548,1550 
and Mg\thinspace II $\lambda\lambda$2796,2800 doublets over the entire 
range $0 < z < 4$ (e.g. Meylan 1995). Moreover, quasar absorption can 
be detected via Lyman-$\alpha$ absorption at an H\thinspace I column 
a million times lower than can be seen directly in emission (Rauch
1998). Lyman-$\alpha$ absorbers are as ubiquitous as galaxies and
they effectively trace the potential of the underlying dark matter 
distribution (Hernquist et al. 1996; Miralda-Escud\'{e} et al. 1996).
Finally, given a sufficiently bright background quasar, absorbers 
can be detected with an efficiency that does not depend on redshift.
By contrast, galaxy surveys are inevitably complicated by effects
such as Malmquist bias, cosmological dimming, k-corrections and
surface brightness selection effects.

The detection of a network of QSO absorbers in a volume centered
on a deep pencil-beam survey allows several interesting experiments
in large scale structure. It allows clustering to be detected on 
scales in excess of 10 $h^{-1}_{100}$ Mpc. Individual QSO sightlines
show C\thinspace IV and Mg\thinspace II correlation power on scales up 
to 20 $h^{-1}_{100}$ Mpc (Quashnock \& Vanden Berk 1998), and multiple
sightlines have been used to trace out three-dimensional structures
on even larger scales (Sargent \& Steidel 1987; Dinshaw \& Impey 1996;
Williger et al. 1996). Deep pencil-beam galaxy redshift surveys have 
shown that around half of the galaxies lie in structures with line 
of sight separations of 50-300 $h^{-1}_{100}$ Mpc (Cohen et al. 1996,
1999). In addition, the spatial relationship between quasar absorbers 
and luminous galaxies can be defined; they are expected to have 
different relationships to the underlying mass distribution (Cen 
et al. 1998). 

We have identified a set of QSOs in the direction of the HDF, suitable
for use as background probes of the volume centered on the HDF line of 
sight. Section 2 contains the multicolor photometry and a simple
multicolor QSO selection strategy. Section 3 includes a description 
of subsequent confirming spectroscopy, a catalog of confirmed QSOs
out to $z \approx 3$ in the magnitude range $17 \lesssim B \lesssim 
21$, and a discussion of the completeness of the survey and the 
properties of the confirmed QSOs. Section 4 is a brief summary, 
along with comments on the applications of these potential absorption 
line probes.

\section{PHOTOMETRY AND CANDIDATE SELECTION} 

\subsection{Multicolor Photometry}

The goal of this study was to obtain a grid of absorption probes out 
to a radius from the HDF of one cluster-cluster correlation scale 
length, or about 8 $h^{-1}_{100}$ Mpc, at the median redshift of the 
deep galaxy surveys, $z = 0.8$. For the cosmology we adopt with $H_0$ 
= 100 km s$^{-1}$ Mpc$^{-1}$ and $q_0$=0.5, this corresponds to 
approximately 30 arcminutes. Our search area was thus the square 
degree centered on the HDF (B1950.0 12:34:35.5 +62:29:28).

Our first broad-band images of the area centered on the HDF were 
obtained in March to May 1996, using the Steward Observatory 2.3-meter 
telescope on Kitt Peak and the Whipple Observatory 1.5-meter telescope 
on Mt. Hopkins. Poor weather for almost all the observing time on these 
runs limited the value of these data. Nevertheless, useful $U$, $B$, 
and $R$ band photometry was obtained for 0.12 square degree centered 
on the HDF. This photometry provided the first QSO candidates for 
spectroscopic followup, which was conducted April 9-10, 1997 at the
Mutiple Mirror Telescope on Mt. Hopkins.

Photometry of the complete square degree was ultimately achieved with
the KPNO 0.9-meter telescope from April 29 to May 5, 1997. We obtained 
$UBR$ photometry of the survey area using the T2KA 2048$\times$2048 
CCD. With a 23$\arcmin$ field of view, the entire square degree was 
covered with a 3$\times$3 mosaic of exposures. Integration times were 
60 minutes in the $U$ band, 60 minutes in the $B$ band, and 30 minutes 
in the $R$ band, each divided into three exposures to facilitate cosmic 
ray rejection. The seeing for these observations ranged from 1.1 to 2.5
arcseconds FWHM. Images were bias-subtracted, flat-fielded and cleaned 
of cosmic rays using the standard routines in IRAF. Objects in the 
reduced images were detected using the Faint Object Classification 
and Analysis System (FOCAS; Valdes 1982). 

Next we used the {\it apphot} package in IRAF to measure aperture photometry 
of each object, using a fixed circular aperture 12 pixels (8$\arcsec$.2) 
in diameter and the sky value sampled from an annulus around each object 
with inner and outer diameters of 16 and 24 pixels respectively. Images 
in the three filters were registered, and positions were measured using 
the COORDS task in IRAF, with Hubble Space Telescope guide stars in 
the frames as a reference grid. The internal rms residuals of the
astrometric solutions ranged from $0\arcsec.2$ to $0\arcsec.5$, which
means there was no ambiguity in comparing objects between filters at
the magnitude limit of this survey. The computed positions of stars in
overlapping regions of the CCD fields matched to within $0\arcsec.5$ 
in all cases.

As with the original imaging observations in the spring of 1996, many 
of the 0.9-meter observations in 1997 were obtained in non-photometric
conditions. Fortunately, enough good weather was available to obtain
flux calibrate each of the $U$, $B$ and $R$ mosaics under photometric
conditions. Absolute photometric calibration was achieved by observing 
standard stars in the globular clusters NGC\thinspace 4147 and 
M\thinspace 92, which were reduced in the same way as the survey data 
described above. We used the {\it photcal} package in IRAF to fit zero points,
color terms, and extinction coefficients. The photometric solutions
yielded rms errors of 0.02, 0.04 and 0.03 magnitudes respectively for 
the $U$, $B$, and $R$ bands. The $10\sigma$ limiting magnitudes for 
point sources were $U = 21.6$, $B = 22.1$ and $R = 21.8$. In all bands, 
point sources brighter than $\sim 15.5$ mag saturated the CCD readout;
this was the practical brightness limit for our photometry.

\subsection{Candidate Selection}

Multicolor selection of QSO candidates is a well understood and widely 
used technique (e.g., Koo, Kron \& Cudworth 1986; Warren et al. 1991; 
Hall et al. 1996). Essentially, the power law energy distributions
of QSOs cause them to be displaced from the stellar locus defined
primarily by hot main sequence stars and white dwarfs. For this work, 
we used a straightforward application of the multicolor selection 
technique based on $U-B$ and $B-R$ colors. The $U-B$ color provides 
optimal sensitivity to ultraviolet-excess QSOs at $z < 2$, while the
$B-R$ color allows the detection of the rarer objects at high redshift.

Figure 1 shows the $U-B$ vs. $B-R$ color-color diagram for the stellar 
objects in the survey area. For clarity, we have included in Figure 1 
only objects with $B < 21.0$ and present half the error bars. The great 
majority of objects lie along the stellar locus, which runs from the 
upper left at ($U-B\sim -0.3, B-R\sim 0.7$) to ($U-B\sim 1.5, B-R\sim 
2.5$). The outliers blueward of the stellar locus in both $U-B$ and 
$B-R$ are most likely to be QSOs. We chose the boundaries for our QSO 
candidates based on both visual inspection of Figure 1, which clearly 
shows the edge of the bulk of the stellar locus, and color-color 
regions used by similar surveys in the literature. Following the work 
of Hall et al. (1996), we adopted a two-stage color selection process
as follows: we consider QSO candidates to be (1) all objects bluer than
$B-R=0.8$, and (2) all objects with both $U-B\leq -0.4$ and $B-R\leq 
1.1$. The boundary in color space of the candidate selection region 
is represented in Figure 1 by the bent solid line.

Given our two-tiered
color-color selection strategy, it is natural to divide our
candidate selection region into three rectangular sub-regions.
As shown in Figure 1, the area ``Q1'' is bounded by $U-B < -0.4$ 
and $B-R < 0.8$, and contains most of the candidates.
Areas ``Q2'' ($U-B > -0.4$ and $B-R < 0.8$) and ``Q3''
($U-B > -0.4$ and $0.8 < B-R < 1.1$) together contain about half
the number of candidates in ``Q1.''

\section{QUASARS IN THE DIRECTION OF THE HDF} 

\subsection{Spectroscopy of QSO Candidates}

Slit spectroscopy of the QSO candidates was obtained with the Multiple
Mirror Telescope between April 1997 and February 1998. Depending on 
the observing run, either the Blue Channel Spectrograph (3200-8000 \AA\
coverage, 6 \AA\ resolution) or the Red Channel Spectrograph (3700-7400 \AA\ 
coverage, 10 \AA\ resolution) was used with a 300 l mm$^{-1}$ grating. As with
the photometry, the data were reduced using the standard methods in the 
{\it ccdred} and {\it longslit} packages in IRAF. Although not all the nights 
were photometric, relative spectrophotometry was obtained for all spectra
using the spectrophotometric standards in Massey \& Gronwall (1990).

Since our primary scientific goal was to find QSOs bright enough to
serve as background probes, we chose a practical limit of $B \sim 21$,
corresponding roughly to the faintest QSO that can be measured at high
resolution within a few hours using the largest ground-based telescopes. 
We thus began by observing all candidates brighter than $B = 21.2$ within 
a 10$\arcmin$ radius of the HDF. Then we observed the brightest candidates 
in the entire square degree, moving progressively fainter as observing 
time and conditions allowed. Over the course of some 14 partial nights, 
with widely varying conditions of transparency and seeing, we observed
a total of 61 candidates in the two-color region described above. This 
number comprises all the stellar objects in that region within the 
survey area from $16 \leq B \leq 20.5$, several fainter targets, and 
all such targets with $B\leq 21.1$ within 10$\arcmin$ of the HDF.

We chose restrictive boundaries for the two-color selection in order 
to maximize the efficiency in the QSO selection and to produce a 
relatively complete spectroscopic sample. But the work of Hall et al. 
(1996), Kennefick et al. (1997) and others have shown that in a UV-excess 
color plot such as one we use, the region near the end of the stellar 
locus, though more strongly contaminated by blue stars, can potentially 
yield additional high-redshift ($z > 2.5$) QSOs. In the hope of confirming
even a few such high-redshift QSOs, we obtained spectra of an additional 
29 randomly selected objects that were redward of the outlier boundary 
we established -- that is, in the approximate color ranges $-0.3\leq 
(U-B)<-0.4$ and $0.8\leq (B-R)<1.0$. Unfortunately, none of these 
borderline candidates were found to be QSOs.

\subsection{Confirmed QSOs}

Our search netted a total of 30 QSOs and 1 AGN.
We present the QSO positions, magnitudes, colors and 
redshifts in Table 1, and their flux-calibrated
spectra in Figure 2. The closest object is the AGN,
a Seyfert galaxy at $z  =0.135$;
all the others have redshift $z = 0.44$ or greater. The most distant QSO 
we identified lies at $z = 2.98$. All of these QSO identifications were 
based either on two or more emission lines, or at least one strong, broad
emission line which we assumed to be MgII at 2800 \AA\ (where any other
choice would have implied another strong line in our spectral window).  
Since the spectra were all flux-calibrated with relative spectrophotometry,
we could also confirm through continuum fitting that the spectra were
consistent with a power law energy distribution. Figure 3 shows the
approximate positions on the sky of the confirmed QSOs with 
respect to the HDF and its flanking fields.

In addition, we present in Table 2 the results of the spectroscopy of the
objects in the color-color regions Q1, Q2 and Q3 (see Figure 1)
that did not yield positive QSO confirmations, 
along with their classification as stars, compact galaxies, or unidentified
sources.  Together, the objects in Tables 1 and 2 comprise all the objects
to the left of the bent solid line in Figure 1 that we have observed.
For reference, we include the spectra of the four unidentified
sources at the end of Figure 2.  

Finally, in Table 3 we 
list the fainter QSO candidates which fall 
into the outlier region for which we did not obtain spectra, down to a 
magnitude limit of $B = 22$. The yield of QSOs within this faint list
could potentially double the QSO sample presented in this paper. These 
tables will hopefully be useful for any future spectroscopic followup 
efforts within the area covered by this survey. 

Looking more closely at the distribution of the 
QSO candidates in color-color space, we find that region Q1
contains 28 of the 30 confirmed QSOs and a 67\% fraction of QSOs to
candidates.  Region Q2 contains 1 AGN and 1 QSO out of 8 candidates,
for a 25\% fraction of active nuclei; both
are relatively low-redshift objects, at $z=$0.135 and 0.58 respectively.  
Region Q3 contains
only one QSO and a 9\% fraction; but that object has the
highest redshift in the sample at z=2.98.
We have listed in Table 3 the color-color region of each faint 
candidate, as a possible indication of how likely the candidate is
to be a QSO.

\subsection{Completeness and QSO Surface Density}

As discussed in Section 3 above, every candidate 
from $16.0\leq B \leq 20.5$ in the regions Q1, Q2 and Q3 
was observed spectroscopically. This totaled 53 objects, 26 of 
which are confirmed as QSOs. Among the 8 fainter candidates observed, 
4 are confirmed as QSOs. So in both subsamples and as a whole, the
selection efficiency is about 50\%. This fraction matches the 46\% 
efficiency achieved by Kennefick et al. (1997) in the magnitude range
$16.5 < B < 21.0$, using very similar color criteria with their $UBV$ 
data.

There are 21 candidates with $20.6\leq B \leq 21.0$ fitting our color 
criteria for which we have no spectra. If we assume that our observational
efficiency is well represented by the 4 out of 8 faint candidates that
are confirmed as QSOs, the entire square degree of this survey should 
contain a total of $\sim 41\pm 6$ QSOs in this magnitude range. This 
prediction is entirely consistent with the observations of Kennefick 
et al. (1997), and with predictions from the results of Koo \& Kron 
(1988) and Boyle, Shanks \& Peterson (1988). The fractions of $z < 2.3$ 
and $z > 2.3$ QSOs that we observe are 27/30 and 3/30, respectively,
which is also consistent at the $1\sigma$ level with the above authors.  
This work is not intended as a study of the quasar luminosity function, 
nor is completeness required to use these QSOs as absorption probes.
However, the consistency of our numbers with those in the literature
means that this catalog fairly represents the QSO population in the 
survey area, and that it has not omitted a large fraction of the QSOs 
in our magnitude range.

\section{SUMMARY} 

We have surveyed the square degree centered on the Hubble Deep Field
for QSOs which can be used as absorption probes, using a straightforward 
optical multi-color selection technique. We present the results of our
spectroscopic identifications, which include 30 confirmed QSOs and 1 AGN
in the magnitude range $17.6 < B < 21.0$ and the redshift 
range $0.14 < z < 2.98$. We also include a list of quasar 
candidates for which spectroscopy 
has not yet been obtained. It is our hope that this work will serve as a
starting point for the establishment of a detailed grid of absorption
probes, in order to study the non-luminous matter within Hubble Deep 
Field volume and its relationship to the galaxy distribution.

\acknowledgments

This work was supported by a NASA archival grant for the HST (AR-06337)
to the University of Arizona. We thank Paul Hewett for
insights into the quasar-hunting business. Additionally, CTL gratefully
acknowledges support from NSF grant AST96-17177 to Columbia University.

\clearpage

\begin{deluxetable}{rlccccccl}
\scriptsize
\tablewidth{0pc}
\tablecaption{Spectroscopic Results -- QSOs}
\tablehead{
\colhead{Number}            & \colhead{Object ID}     &
\colhead{R.A. (1950)}       & \colhead{Dec. (1950)}    &
\colhead{B}             & \colhead{U-B}           &
\colhead{B-R}           & \colhead{$z$}           &
\colhead{Notes}           
}

\startdata

 1 & Q1230+6226  & 12 30 12.9 & 62 26 23 & 19.4 & -0.95 &  0.41 & 2.07 
   & Ly$\alpha$,C\thinspace IV,C\thinspace III \nl %
 2 & Q1230+6215  & 12 30 43.6 & 62 15 34 & 19.4 & -0.84 &  0.71 & 1.47 
   & C\thinspace IV,C\thinspace III \nl %
 3 & Q1230+6225  & 12 30 47.6 & 62 25 49 & 18.9 & -0.59 &  0.62 & 1.83 
   & Si\thinspace IV/O\thinspace VI, C\thinspace IV, C\thinspace III \nl %
 4 & Q1230+6249  & 12 30 59.7 & 62 49 44 & 20.5 & -0.53 &  0.70 & 0.80 
   & C\thinspace III, Mg\thinspace II \nl %
 5 & Q1231+6249  & 12 31 17.5 & 62 49 26 & 20.4 & -0.87 &  0.70 & 1.32 
   & C\thinspace IV,C\thinspace III \nl %
 6 & Q1231+6227  & 12 31 23.8 & 62 27 39 & 20.3 & -0.68 &  0.71 & 0.50 
   & Mg\thinspace II  \nl %
 7 & Q1231+6249  & 12 31 45.9 & 62 49 47 & 19.9 & -0.82 &  0.59 & 1.12 
   & C\thinspace III, Mg\thinspace II \nl %
 8 & Q1231+6244  & 12 31 47.3 & 62 44 24 & 19.5 & -0.79 &  0.78 & 1.31 
   & C\thinspace III, Mg\thinspace II \nl %
 9 & Q1231+6215  & 12 31 56.3 & 62 15 05 & 19.1 & -0.68 &  0.77 & 1.95 
   & Ly$\alpha$,C\thinspace IV,C\thinspace III \nl %
10 & Q1231+6243  & 12 31 59.7 & 62 43 12 & 17.6 & -0.88 &  0.70 & 1.33 
   & C\thinspace III, Mg\thinspace II \nl %
11 & Q1232+6207  & 12 32 57.9 & 62 07 28 & 20.6 & -0.66 &  0.57 & 0.98 
   & C\thinspace III, Mg\thinspace II \nl %
12 & Q1233+6221  & 12 33 55.9 & 62 21 06 & 20.5 & -0.79 &  0.68 & 1.74 
   & Ly$\alpha$,C\thinspace IV \nl %
13 & Q1234+6231  & 12 34 08.8 & 62 31 58 & 21.0 & -0.43 &  0.66 & 2.58 
   & Ly$\alpha$,C\thinspace IV,C\thinspace III \nl %
14 & Q1234+6214  & 12 34 23.3 & 62 14 45 & 19.4 & -0.43 &  0.73 & 2.52 
   & Ly$\alpha$,C\thinspace IV,C\thinspace III \nl %
15 & Q1235+6219  & 12 35 02.3 & 62 19 54 & 20.3 & -0.92 &  0.67 & 2.05 
   & Ly$\alpha$,C\thinspace IV,C\thinspace III \nl %
16 & Q1235+6205  & 12 35 33.7 & 62 05 48 & 19.1 & -0.61 &  0.64 & 2.28 
   & Ly$\alpha$,C\thinspace IV \nl %
17 & Q1235+6230  & 12 35 47.6 & 62 30 06 & 19.2 & -0.80 &  0.74 & 0.44 
   & Mg\thinspace II,H$\beta$ \nl %
18 & Q1235+6243  & 12 35 59.2 & 62 43 56 & 20.8 & -0.49 &  0.63 & 0.77 
   & C\thinspace III, Mg\thinspace II \nl %
19 & Q1236+6241  & 12 36 02.7 & 62 41 08 & 20.9 & -0.66 &  0.69 & 1.75 
   & Ly$\alpha$,C\thinspace IV \nl %
20 & Q1236+6218  & 12 36 02.9 & 62 18 38 & 19.3 & -0.64 &  0.58 & 1.00
   & C\thinspace III, Mg\thinspace II \nl %
21 & Q1236+6203  & 12 36 10.6 & 62 03 42 & 20.0 & -0.28 &  0.79 & 2.98 
   & Ly$\alpha$,C\thinspace IV \nl %
22 & Q12364+6200 & 12 36 21.5 & 62 00 38 & 20.1 & -0.61 &  0.57 & 2.27 
   & Ly$\alpha$,C\thinspace IV,C\thinspace III  \nl %
23 & Q12366+6200 & 12 36 35.4 & 62 00 23 & 18.2 & -0.60 &  0.61 & 0.83 
   & C\thinspace III, Mg\thinspace II \nl %
24 & Q1236+6158  & 12 36 49.2 & 61 58 39 & 20.3 & -0.47 &  0.70 & 0.91 
   & Mg\thinspace II \nl %
25 & Q1237+6222  & 12 37 19.0 & 62 22 47 & 19.6 & -0.65 &  0.68 & 1.19 
   & C\thinspace IV,C\thinspace III \nl %
26 & Q1237+6249  & 12 37 22.2 & 62 49 47 & 20.1 & -0.56 &  0.72 & 1.66 
   & Ly$\alpha$,C\thinspace IV,C\thinspace III \nl %
27 & Q1238+6239  & 12 38 21.5 & 62 39 56 & 20.5 & -0.58 &  0.70 & 1.09 
   & C\thinspace III, Mg\thinspace II \nl %
28 & Q1238+6232  & 12 38 22.3 & 62 32 33 & 20.4 & -0.59 &  0.98 & 0.58 
   & Mg\thinspace II,O\thinspace II, H$\delta$ \nl %
29 & Q1238+6252  & 12 38 36.8 & 62 52 21 & 19.3 & -0.66 &  0.55 & 1.78 
   & Ly$\alpha$,C\thinspace IV \nl %
30 & Q1238+6205  & 12 38 45.9 & 62 05 03 & 20.1 & -0.71 &  0.58 & 1.08
   & C\thinspace III, Mg\thinspace II \nl %

\enddata
\end{deluxetable}

\begin{deluxetable}{llcccl}
\scriptsize
\tablewidth{0pc}
\tablecaption{Spectroscopic Results -- non-QSOs}
\tablehead{
\colhead{R.A. (1950)}       & \colhead{Dec. (1950)}    &
\colhead{B}             & \colhead{U-B}           &
\colhead{B-R}               & \colhead{ID}           
}

\startdata

12 32 14.8 & 62 34 38 & 17.8 & -0.41 & 0.84 & AGN, z=0.135    \nl
12 36 36.1 & 62 08 45 & 19.6 & -0.32 & 0.79 & ?      \nl 
12 32 13.2 & 62 23 03 & 20.5 & -0.48 & 0.81 & ?      \nl
12 33 14.2 & 62 49 00 & 20.8 & -0.68 & 0.68 & ?      \nl
12 38 39.4 & 62 53 01 & 20.9 & -0.57 & 0.76 & ?      \nl
12 35 59.0 & 62 39 08 & 20.4 & -0.41 & 0.80 & galaxy, z=0.232 \nl
12 36 58.1 & 62 18 46 & 16.8 & -1.13 & 0.09 & star   \nl
12 33 47.9 & 62 05 06 & 17.1 & -0.35 & 0.58 & star   \nl
12 37 41.3 & 62 24 23 & 18.2 & -0.37 & 0.77 & star   \nl
12 38 38.9 & 62 52 15 & 18.4 & -0.65 & 0.31 & star   \nl
12 31 26.6 & 62 24 30 & 18.9 & -0.27 & 0.78 & star   \nl
12 35 35.5 & 62 00 13 & 19.1 & -0.48 & 0.91 & star   \nl
12 37 39.6 & 62 07 24 & 19.4 & -0.89 & 0.34 & star   \nl
12 34 27.8 & 62 52 48 & 19.5 & -0.31 & 0.79 & star   \nl
12 37 41.9 & 62 16 24 & 19.6 & -0.32 & 0.75 & star   \nl
12 33 30.1 & 62 50 45 & 19.7 & -1.04 & 0.22 & star   \nl
12 34 11.5 & 62 50 53 & 19.8 & -0.43 & 0.58 & star   \nl
12 36 07.7 & 62 35 09 & 20.0 & -0.09 & 0.65 & star   \nl
12 35 39.4 & 62 32 58 & 20.0 & -0.52 & 1.03 & star   \nl
12 32 49.5 & 62 21 50 & 20.0 & -0.79 & 0.45 & star   \nl
12 30 44.8 & 62 19 40 & 20.0 & -0.41 & 0.61 & star   \nl
12 32 08.7 & 62 33 07 & 20.2 & -0.36 & 0.75 & star   \nl
12 36 09.4 & 62 55 57 & 20.3 & -1.06 & 0.17 & star   \nl
12 30 04.1 & 62 38 13 & 20.4 & -0.34 & 0.75 & star   \nl
12 33 48.7 & 62 30 08 & 20.4 & -0.45 & 0.81 & star   \nl
12 36 40.5 & 62 53 13 & 20.4 & -0.42 & 0.64 & star   \nl
12 35 51.2 & 62 26 45 & 20.5 & -0.38 & 0.68 & star   \nl
12 33 09.7 & 62 06 28 & 20.5 & -0.67 & 0.52 & star   \nl
12 32 55.1 & 62 49 25 & 20.5 & -0.48 & 0.83 & star   \nl
12 38 22.3 & 62 51 24 & 20.9 & -0.48 & 0.73 & star   \nl
12 34 31.5 & 62 28 44 & 21.0 & -0.92 & 0.34 & star   \nl

\enddata
\end{deluxetable}

\begin{deluxetable}{rlcccc}
\scriptsize
\tablewidth{0pc}
\tablecaption{QSO candidates}
\tablehead{
\colhead{R.A. (1950)}   & \colhead{Dec. (1950)}    &
\colhead{B}             & \colhead{U-B}           &
\colhead{B-R}           & \colhead{Region}    
}

\startdata

12 31 36.0 & 63 00 21 & 20.6 & -0.43 & 0.99 & Q3  \nl
12 30 53.4 & 62 04 30 & 20.6 & -0.77 & 0.60 & Q1  \nl
12 30 53.5 & 62 17 35 & 20.7 & -0.85 & 0.54 & Q1  \nl
12 30 45.1 & 62 48 51 & 20.7 & -0.57 & 0.73 & Q1  \nl
12 38 19.8 & 62 18 34 & 20.8 & -0.06 & 0.76 & Q2  \nl
12 37 17.7 & 62 11 02 & 20.8 & -0.51 & 0.94 & Q3  \nl
12 37 56.6 & 62 18 48 & 20.8 & -0.40 & 0.77 & Q2  \nl
12 38 53.6 & 62 23 15 & 20.9 & -0.44 & 1.06 & Q3  \nl
12 33 17.7 & 62 30 01 & 20.9 & -0.34 & 0.67 & Q2  \nl
12 35 48.4 & 62 36 50 & 20.9 & -0.47 & 0.74 & Q1  \nl
12 37 14.4 & 62 50 32 & 20.9 & -0.88 & 0.73 & Q1  \nl
12 34 53.2 & 62 42 20 & 20.9 & -0.58 & 1.01 & Q3  \nl
12 38 56.9 & 61 58 32 & 20.9 & -0.58 & 0.76 & Q1  \nl
12 37 49.7 & 62 25 07 & 21.0 & -0.49 & 0.68 & Q1  \nl
12 38 58.9 & 62 21 59 & 21.0 & -0.18 & 0.67 & Q2  \nl
12 32 24.4 & 62 24 00 & 21.0 & -0.64 & 0.66 & Q1  \nl
12 35 18.1 & 62 39 36 & 21.0 & -0.46 & 0.83 & Q3  \nl
12 30 33.7 & 62 14 17 & 21.0 & -0.52 & 0.61 & Q1  \nl
12 31 18.7 & 62 25 36 & 21.0 & -0.30 & 0.75 & Q2  \nl
12 35 17.2 & 62 18 51 & 21.0 & -0.63 & 0.74 & Q1  \nl
12 36 35.1 & 62 30 55 & 21.0 & -0.68 & 0.48 & Q1  \nl
12 32 26.6 & 62 43 09 & 21.1 & -0.42 & 0.75 & Q1  \nl
12 34 27.6 & 62 21 12 & 21.1 & -0.44 & 0.80 & Q3  \nl
12 31 41.0 & 62 29 58 & 21.1 & -0.31 & 0.66 & Q2  \nl
12 33 48.2 & 62 50 05 & 21.1 & -0.53 & 0.88 & Q3  \nl
12 31 22.8 & 62 10 19 & 21.1 & -1.06 & 0.57 & Q1  \nl
12 37 24.3 & 62 22 01 & 21.1 & -0.47 & 0.89 & Q3  \nl
12 30 41.4 & 62 32 18 & 21.1 & -0.61 & 0.66 & Q1  \nl
12 37 07.5 & 62 00 35 & 21.1 & -0.45 & 0.77 & Q1  \nl
12 35 51.8 & 62 45 59 & 21.1 & -0.88 & 0.55 & Q1  \nl
12 33 56.1 & 62 33 01 & 21.2 & -0.62 & 0.80 & Q3  \nl
12 32 18.1 & 62 31 46 & 21.2 & -0.65 & 0.62 & Q1  \nl
12 31 21.9 & 62 52 44 & 21.2 & -0.70 & 1.03 & Q3  \nl
12 36 40.0 & 62 19 07 & 21.2 & -0.34 & 0.76 & Q2  \nl
12 34 35.9 & 62 46 04 & 21.2 & -0.37 & 0.62 & Q2  \nl
12 31 40.1 & 62 44 34 & 21.2 & -0.44 & 0.82 & Q3  \nl
12 31 42.7 & 62 56 23 & 21.2 & -0.10 & 0.73 & Q2  \nl
12 30 23.0 & 62 27 23 & 21.2 & -0.96 & 0.64 & Q1  \nl
12 32 50.8 & 62 09 15 & 21.2 & -1.16 & 0.26 & Q1  \nl
12 31 21.6 & 62 26 59 & 21.2 & -0.47 & 0.77 & Q1  \nl
12 36 55.7 & 62 33 39 & 21.2 & -0.56 & 0.86 & Q3  \nl
12 31 42.1 & 62 16 04 & 21.3 & -0.82 & 1.01 & Q3  \nl
12 35 10.0 & 62 58 34 & 21.3 & -0.53 & 1.08 & Q3  \nl
12 38 58.9 & 62 51 39 & 21.3 & -0.90 & 0.90 & Q3  \nl
12 33 58.0 & 62 36 13 & 21.3 & -0.99 & 0.44 & Q1  \nl
12 35 54.2 & 62 48 21 & 21.3 & -0.43 & 0.97 & Q3  \nl
12 36 26.6 & 62 17 13 & 21.3 & -0.28 & 0.74 & Q2  \nl
12 36 45.1 & 62 23 46 & 21.3 & -0.36 & 0.68 & Q2  \nl
12 31 54.5 & 62 06 43 & 21.3 & -0.49 & 0.75 & Q1  \nl
12 38 17.3 & 62 50 23 & 21.3 & -0.48 & 0.94 & Q3  \nl
12 34 05.9 & 62 47 34 & 21.4 & -0.72 & 0.77 & Q1  \nl
12 38 07.2 & 62 05 09 & 21.4 & -0.24 & 0.62 & Q2  \nl
12 31 25.0 & 62 22 03 & 21.4 & -0.49 & 0.82 & Q3  \nl
12 33 28.5 & 62 25 06 & 21.4 & -0.61 & 0.97 & Q3  \nl
12 35 31.5 & 62 13 42 & 21.4 & -0.43 & 1.07 & Q3  \nl
12 32 13.0 & 62 24 55 & 21.4 & -1.15 & 0.59 & Q1  \nl
12 32 51.3 & 61 59 33 & 21.4 & -0.41 & 1.05 & Q3  \nl
12 35 07.4 & 62 08 03 & 21.4 & -2.11 & -.16 & Q1  \nl
12 32 22.5 & 62 48 54 & 21.4 & -0.53 & 0.91 & Q3  \nl
12 34 23.8 & 62 01 14 & 21.4 & -0.56 & 0.65 & Q1  \nl
12 31 54.7 & 62 23 08 & 21.4 & -0.12 & 0.78 & Q2  \nl
12 38 56.4 & 62 05 04 & 21.4 & -0.45 & 1.03 & Q3  \nl
12 31 17.9 & 62 52 46 & 21.4 & -0.64 & 0.94 & Q3  \nl
12 31 15.7 & 62 41 09 & 21.4 & -0.76 & 0.58 & Q1  \nl
12 35 40.1 & 62 10 15 & 21.5 & -0.28 & 0.76 & Q2  \nl
12 37 19.3 & 62 34 16 & 21.5 & -0.51 & 0.83 & Q3  \nl
12 30 57.6 & 62 09 18 & 21.5 & -0.73 & 0.42 & Q1  \nl
12 30 57.0 & 62 14 06 & 21.5 & -0.33 & 0.76 & Q2  \nl
12 36 11.7 & 62 09 06 & 21.5 & -0.55 & 0.57 & Q1  \nl
12 31 59.8 & 62 11 33 & 21.5 & -0.50 & 0.63 & Q1  \nl
12 33 16.9 & 62 09 06 & 21.5 & -0.68 & 0.90 & Q3  \nl
12 36 03.2 & 62 38 20 & 21.5 & -0.42 & 0.93 & Q3  \nl
12 35 53.1 & 62 29 53 & 21.5 & -0.50 & 0.86 & Q3  \nl
12 36 23.9 & 62 20 26 & 21.5 & -0.63 & 1.01 & Q3  \nl
12 30 34.0 & 62 40 42 & 21.5 & -0.15 & 0.62 & Q2  \nl
12 35 51.9 & 62 04 45 & 21.6 & -0.43 & 0.96 & Q3  \nl
12 30 19.9 & 62 28 37 & 21.6 & -1.00 & 0.10 & Q1  \nl
12 34 51.1 & 62 31 25 & 21.6 & -0.39 & 0.73 & Q2  \nl
12 39 00.9 & 62 29 44 & 21.6 & -0.50 & 0.98 & Q3  \nl
12 37 01.8 & 62 26 25 & 21.6 & -0.56 & 0.86 & Q3  \nl
12 30 56.6 & 62 20 14 & 21.6 & -0.70 & 0.90 & Q3  \nl
12 36 03.2 & 62 46 44 & 21.6 & -0.86 & 0.60 & Q1  \nl
12 34 51.5 & 62 20 33 & 21.6 & -0.47 & 0.79 & Q1  \nl
12 34 53.0 & 62 47 50 & 21.6 & -0.75 & 0.68 & Q1  \nl
12 30 47.9 & 62 11 39 & 21.6 & -0.88 & 0.78 & Q1  \nl
12 34 16.7 & 62 42 44 & 21.6 & -0.33 & 0.73 & Q2  \nl
12 31 29.4 & 62 34 20 & 21.6 & -0.37 & 0.79 & Q2  \nl
12 30 31.7 & 62 28 59 & 21.6 & -0.50 & 0.91 & Q3  \nl
12 30 13.2 & 62 23 02 & 21.6 & -0.42 & 1.08 & Q3  \nl
12 35 02.6 & 62 56 35 & 21.6 & -0.45 & 0.73 & Q1  \nl
12 31 54.6 & 62 40 11 & 21.6 & -0.50 & 0.88 & Q3  \nl
12 31 03.9 & 62 34 42 & 21.6 & -0.53 & 0.89 & Q3  \nl
12 32 52.4 & 62 38 11 & 21.6 & -0.75 & 0.93 & Q3  \nl
12 32 18.4 & 62 09 43 & 21.6 & -0.76 & 0.97 & Q3  \nl
12 35 46.7 & 62 56 26 & 21.6 & -0.72 & 0.89 & Q3  \nl
12 37 07.6 & 62 15 08 & 21.6 & -0.74 & 0.80 & Q3  \nl
12 31 05.3 & 62 03 44 & 21.6 & -1.45 & 0.89 & Q3  \nl
12 34 32.5 & 62 44 24 & 21.7 & -0.44 & 0.87 & Q3  \nl
12 32 55.5 & 62 41 20 & 21.7 & -0.50 & 0.77 & Q1  \nl
12 30 39.4 & 62 47 53 & 21.7 & -0.96 & 0.84 & Q3  \nl
12 35 33.6 & 62 45 25 & 21.7 & -0.44 & 1.09 & Q3  \nl
12 36 25.7 & 62 21 13 & 21.7 & -0.41 & 0.91 & Q3  \nl
12 33 34.4 & 62 39 02 & 21.7 & -0.52 & 0.92 & Q3  \nl
12 30 23.8 & 62 55 47 & 21.7 & -0.48 & 1.01 & Q3  \nl
12 32 27.2 & 62 24 47 & 21.7 & -0.42 & 0.81 & Q3  \nl
12 32 19.7 & 62 33 36 & 21.7 & -0.47 & 0.93 & Q3  \nl
12 35 44.0 & 62 14 05 & 21.7 & -0.59 & 0.71 & Q1  \nl
12 35 53.5 & 61 59 10 & 21.7 & -0.78 & 0.69 & Q1  \nl
12 31 08.0 & 62 45 25 & 21.7 & -0.81 & 0.94 & Q3  \nl
12 36 20.5 & 62 20 21 & 21.7 & -1.05 & 0.46 & Q1  \nl
12 35 47.9 & 62 58 37 & 21.7 & -0.55 & 0.75 & Q1  \nl
12 31 55.4 & 62 31 01 & 21.7 & -0.72 & 0.62 & Q1  \nl
12 35 35.6 & 62 26 41 & 21.8 & -0.54 & 1.04 & Q3  \nl
12 30 24.8 & 62 26 29 & 21.8 & -0.69 & 0.55 & Q1  \nl
12 34 13.3 & 62 26 50 & 21.8 & -0.36 & 0.78 & Q2  \nl
12 34 37.9 & 62 34 06 & 21.8 & -0.46 & 0.81 & Q3  \nl
12 38 08.9 & 62 35 25 & 21.8 & -0.54 & 0.65 & Q1  \nl
12 34 29.6 & 62 38 33 & 21.8 & -0.60 & 0.82 & Q3  \nl
12 31 19.6 & 62 07 59 & 21.8 & -0.93 & 0.45 & Q1  \nl
12 35 54.7 & 62 37 14 & 21.8 & -0.19 & 0.72 & Q2  \nl
12 34 05.1 & 62 07 32 & 21.8 & -0.44 & 0.58 & Q1  \nl
12 30 18.2 & 62 22 10 & 21.8 & -0.44 & 0.95 & Q3  \nl
12 33 40.9 & 62 21 06 & 21.8 & -0.73 & 0.87 & Q3  \nl
12 34 44.5 & 62 52 02 & 21.8 & -0.41 & 0.87 & Q3  \nl
12 36 09.8 & 62 40 31 & 21.8 & -0.41 & 0.91 & Q3  \nl
12 33 44.1 & 62 34 29 & 21.8 & -0.51 & 0.66 & Q1  \nl
12 34 18.5 & 62 41 21 & 21.8 & -0.66 & 1.02 & Q3  \nl
12 37 35.1 & 62 04 30 & 21.8 & -1.64 & 0.35 & Q1  \nl
12 31 01.6 & 62 19 36 & 21.8 & -0.51 & 0.78 & Q1  \nl
12 34 43.8 & 62 45 53 & 21.8 & -1.11 & 0.90 & Q3  \nl
12 34 04.4 & 62 27 46 & 21.8 & -0.92 & 0.72 & Q1  \nl
12 31 58.3 & 62 35 08 & 21.9 & -0.33 & 0.68 & Q2  \nl
12 36 51.1 & 62 21 33 & 21.9 & -0.79 & 0.72 & Q1  \nl
12 35 48.2 & 62 54 22 & 21.9 & -0.89 & 0.51 & Q1  \nl
12 34 34.5 & 62 24 24 & 21.9 & -0.31 & 0.74 & Q2  \nl
12 34 50.5 & 62 02 18 & 21.9 & -0.62 & 0.53 & Q1  \nl
12 31 18.1 & 62 43 23 & 21.9 & -0.54 & 0.80 & Q3  \nl
12 31 55.0 & 62 19 32 & 21.9 & -0.45 & 0.89 & Q3  \nl
12 35 03.2 & 62 22 24 & 21.9 & -0.77 & 0.67 & Q1  \nl
12 36 12.0 & 62 46 07 & 21.9 & -0.61 & 0.98 & Q3  \nl
12 36 06.6 & 62 49 14 & 21.9 & -0.95 & 0.79 & Q1  \nl
12 31 59.5 & 62 23 23 & 21.9 & -0.35 & 0.72 & Q2  \nl
12 31 12.7 & 62 29 48 & 22.0 & -1.01 & 0.80 & Q3  \nl
12 32 57.4 & 62 52 10 & 22.0 & -1.05 & 0.20 & Q1  \nl
12 37 00.7 & 62 09 14 & 22.0 & -1.09 & 0.50 & Q1  \nl
12 36 02.9 & 62 23 00 & 22.0 & -0.48 & 0.98 & Q3  \nl
12 35 23.3 & 62 42 54 & 22.0 & -0.70 & 0.54 & Q1  \nl
12 32 27.9 & 62 48 29 & 22.0 & -0.93 & 1.00 & Q3  \nl
12 32 58.6 & 62 19 31 & 22.0 & -0.48 & 0.98 & Q3  \nl
12 32 07.5 & 62 40 56 & 22.0 & -0.71 & 1.01 & Q3  \nl
12 30 53.6 & 62 30 26 & 22.0 & -0.81 & 0.99 & Q3  \nl

\enddata
\end{deluxetable}

\clearpage

%

\begin{figure}
\epsscale{1}
\plotone{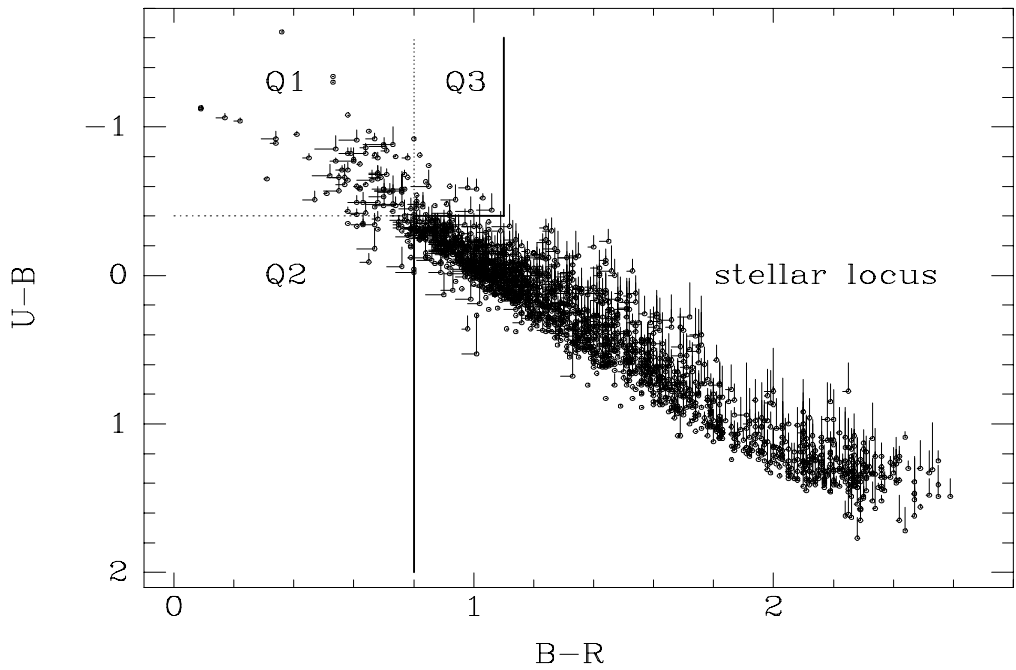}
\caption{$U-B$ vs. $B-R$ diagram for all stellar objects in the 
survey area with $B < 21.0$. For clarity, only half the error bars
are plotted for each data point. The bent solid line denotes the boundary 
that marks the approximate end of the stellar locus; objects bluer
than this have a high probability of being QSOs.  The dotted lines
mark the boundaries of the color-color regions Q1, Q2 and Q3 (see text).
}
\end{figure}
\clearpage

\begin{figure}
\epsscale{1.2}
\plotfiddle{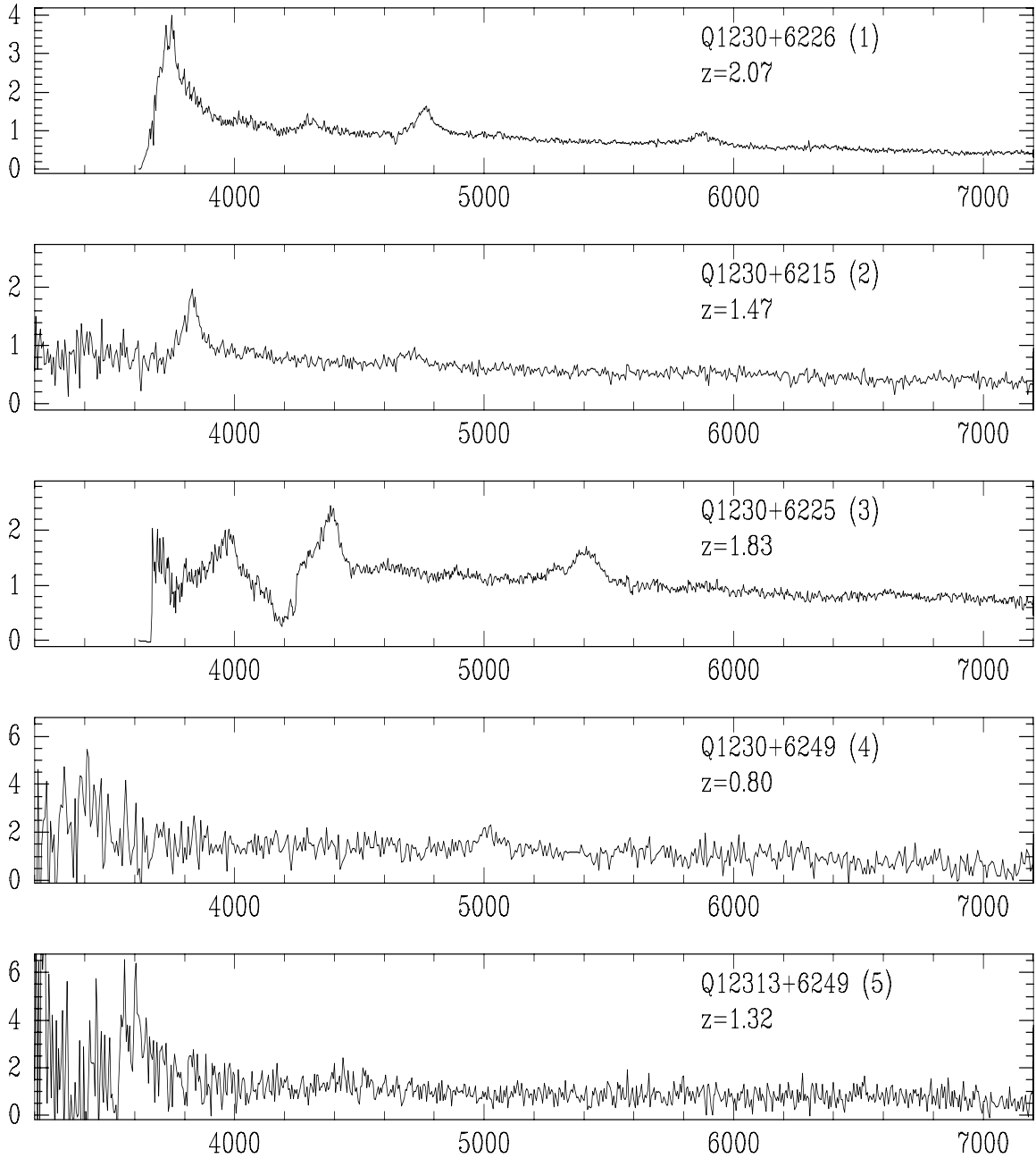}{7in}{0}{100.}{160.}{-324}{-360}
\caption{Spectra of the QSOs identified from the spectroscopy.
Wavelength is in \AA. Flux is presented in units of $10^{-16}$ ergs 
cm$^{-2}$ sec$^{-1}$ \AA$^{-1}$. Redshifts are based on the detection
of two or more broad emission features.  The last four spectra have
an uncertain classification.
}
\end{figure}
\clearpage

\begin{figure}
\epsscale{1.2}
\plotfiddle{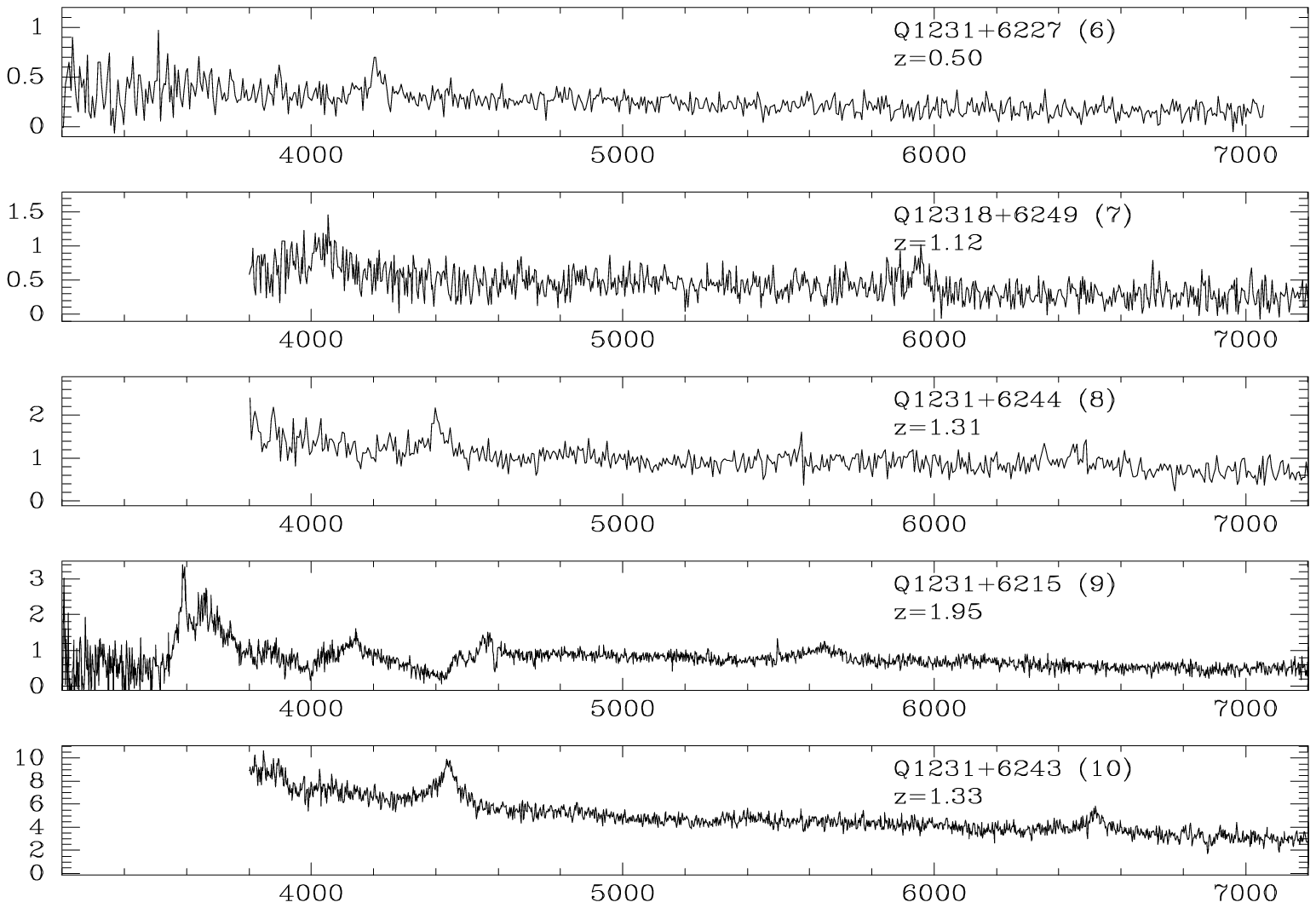}{7in}{0}{100.}{160.}{-324}{-360}
\end{figure}
\clearpage

\begin{figure}
\epsscale{1.2}
\plotfiddle{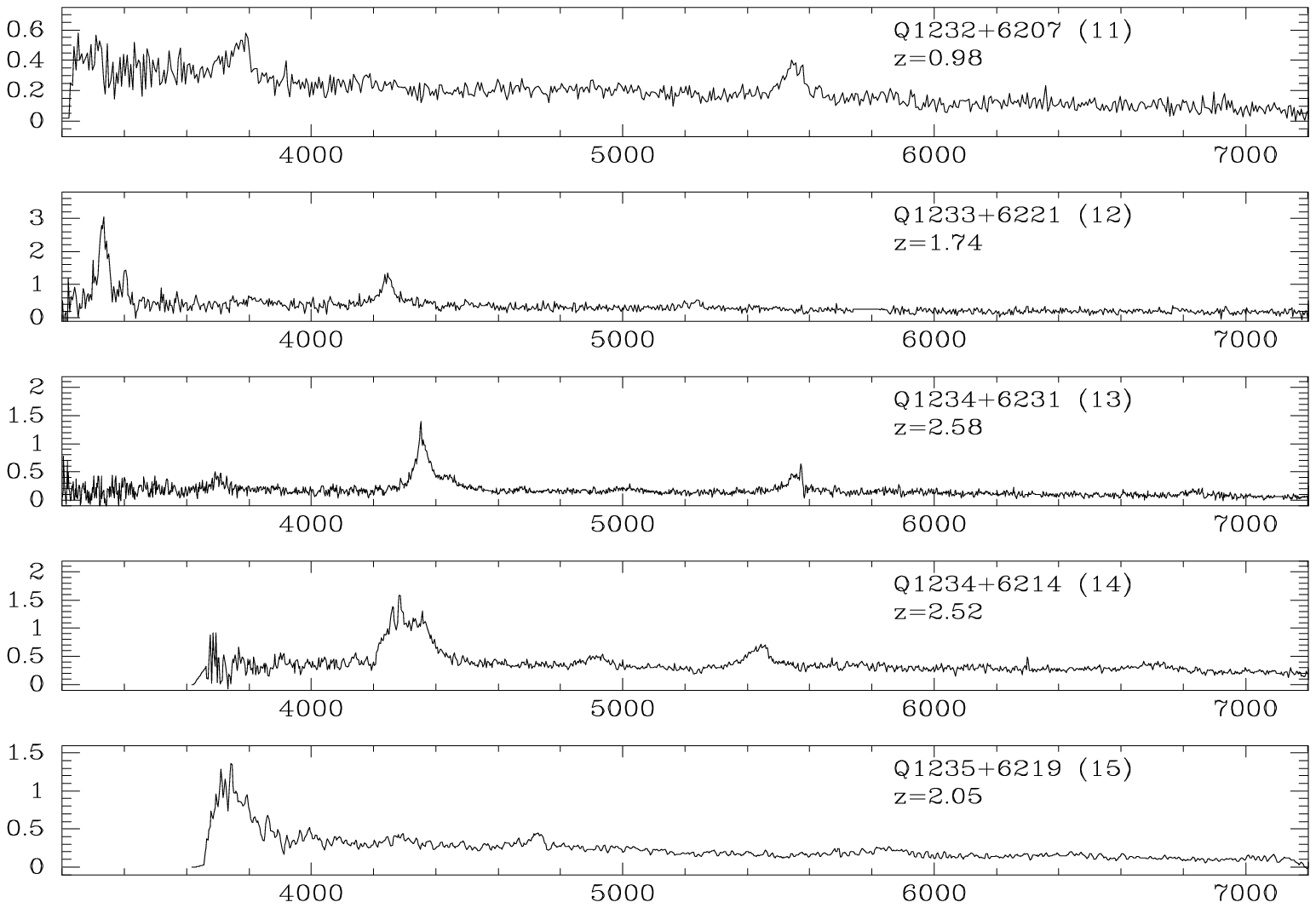}{7in}{0}{100.}{160.}{-324}{-360}
\end{figure}
\clearpage

\begin{figure}
\epsscale{1.2}
\plotfiddle{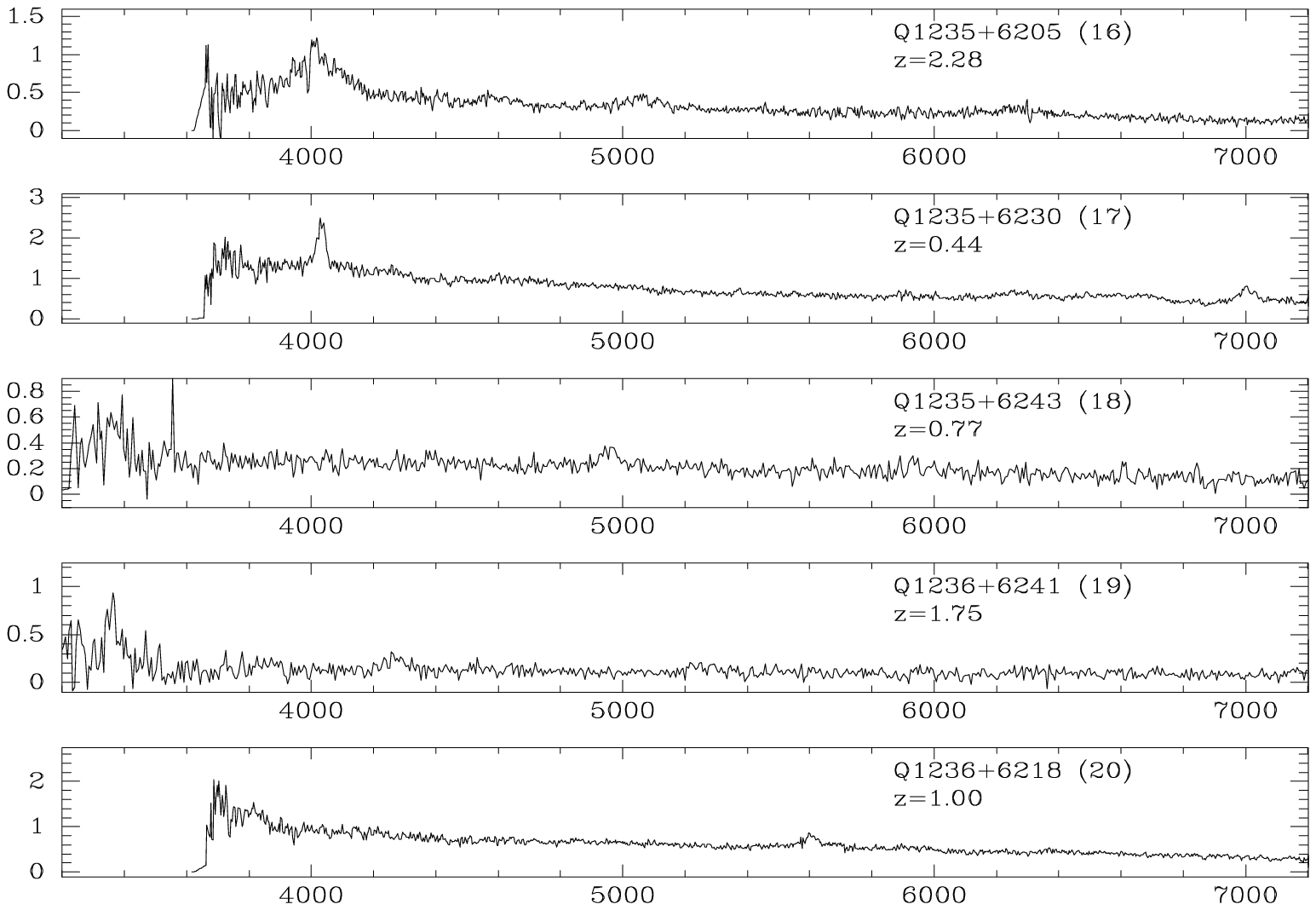}{7in}{0}{100.}{160.}{-324}{-360}
\end{figure}
\clearpage

\begin{figure}
\epsscale{1.2}
\plotfiddle{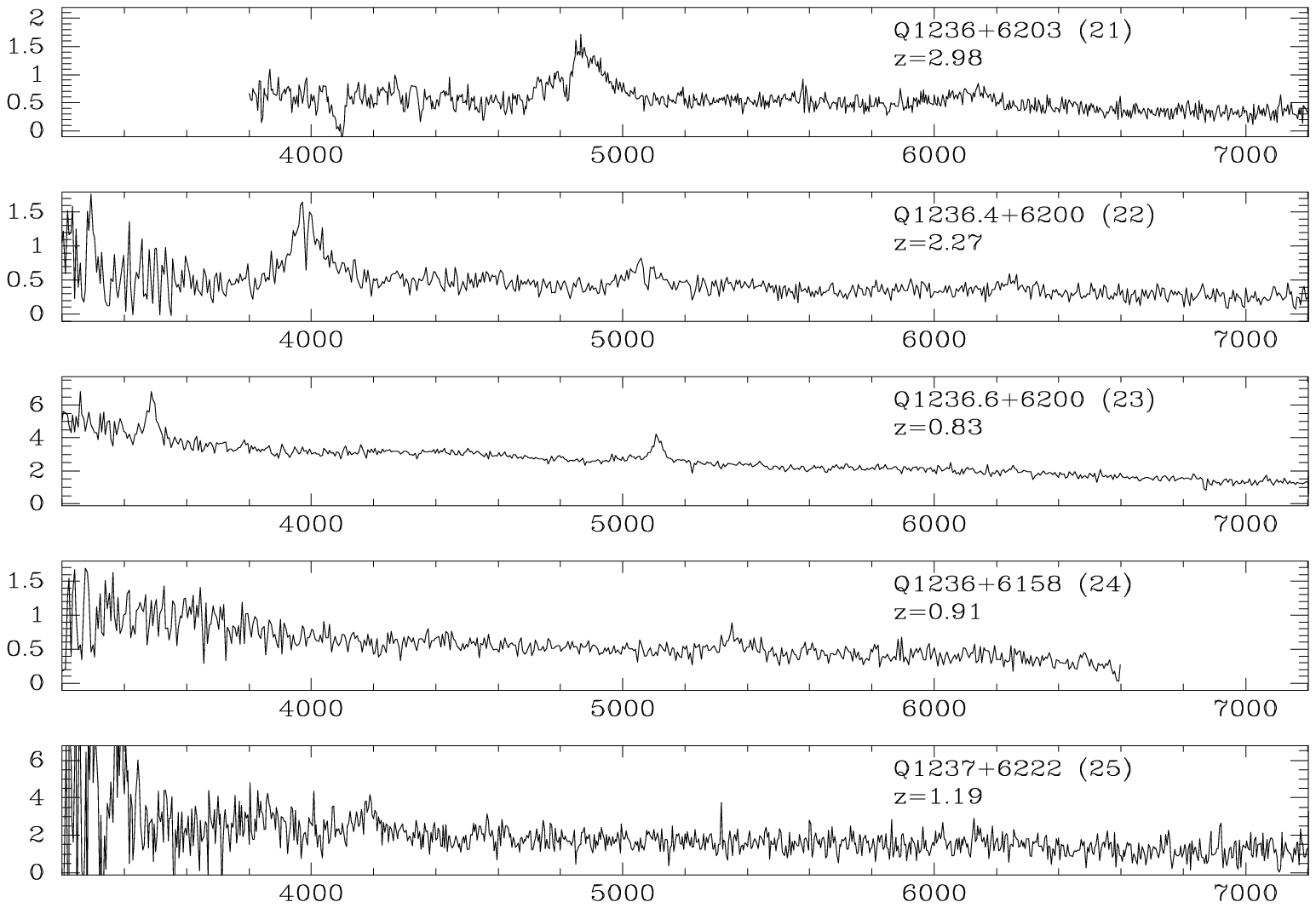}{7in}{0}{100.}{160.}{-324}{-360}
\end{figure}
\clearpage

\begin{figure}
\epsscale{1.2}
\plotfiddle{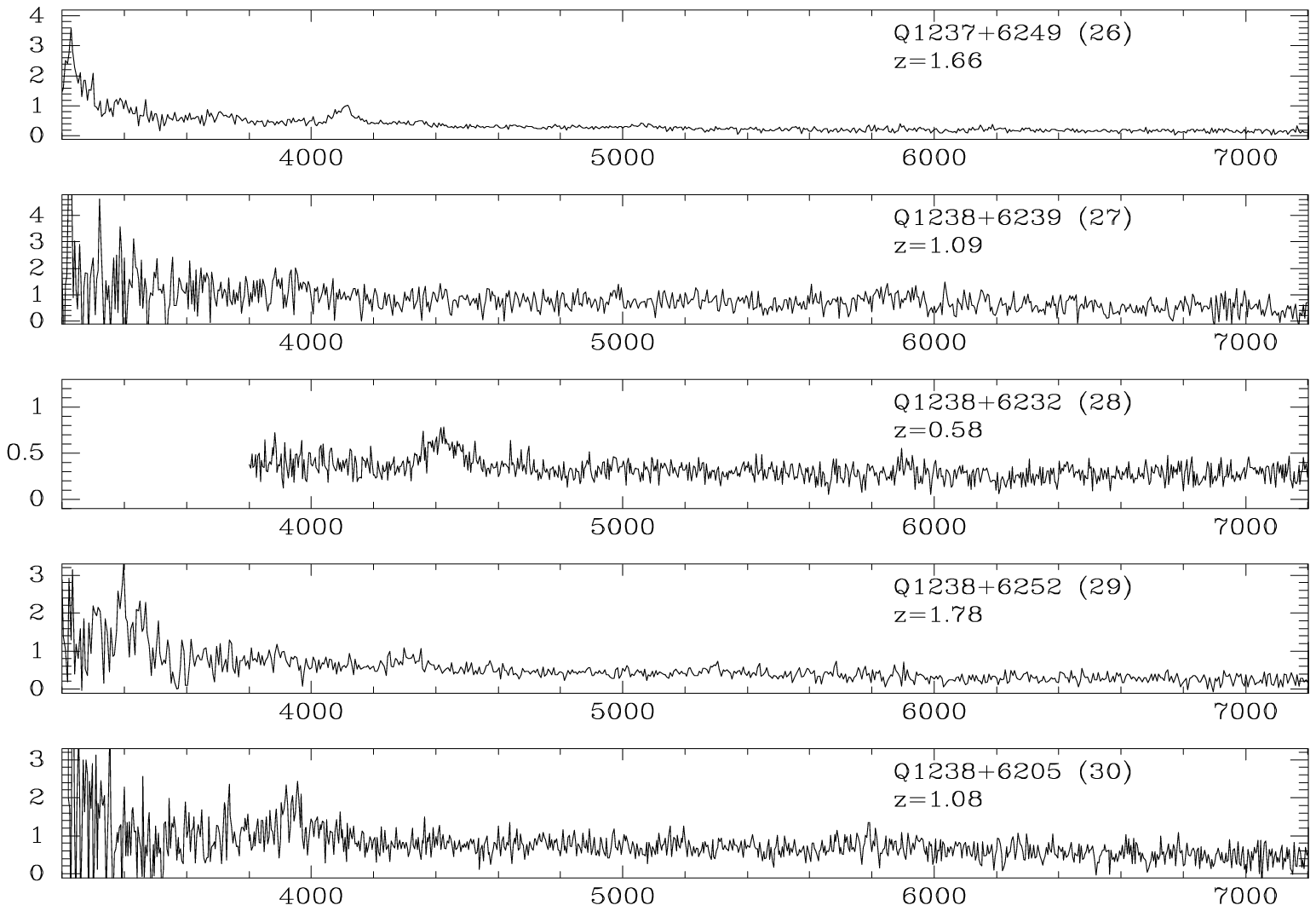}{7in}{0}{100.}{160.}{-324}{-360}
\end{figure}
\clearpage

\begin{figure}
\epsscale{1.2}
\plotfiddle{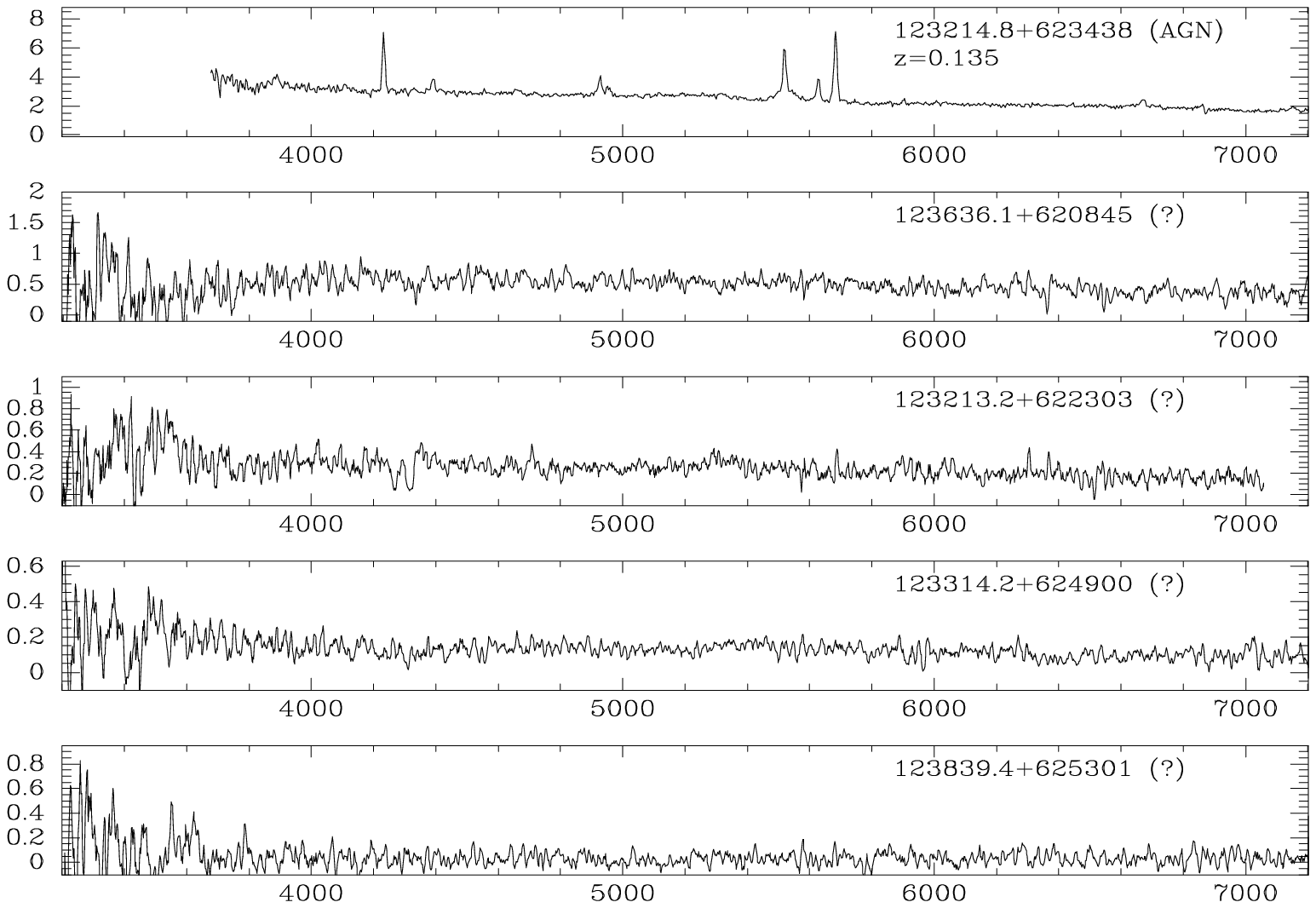}{7in}{0}{100.}{160.}{-324}{-360}
\end{figure}
\clearpage

\begin{figure}
\epsscale{1}
\plotfiddle{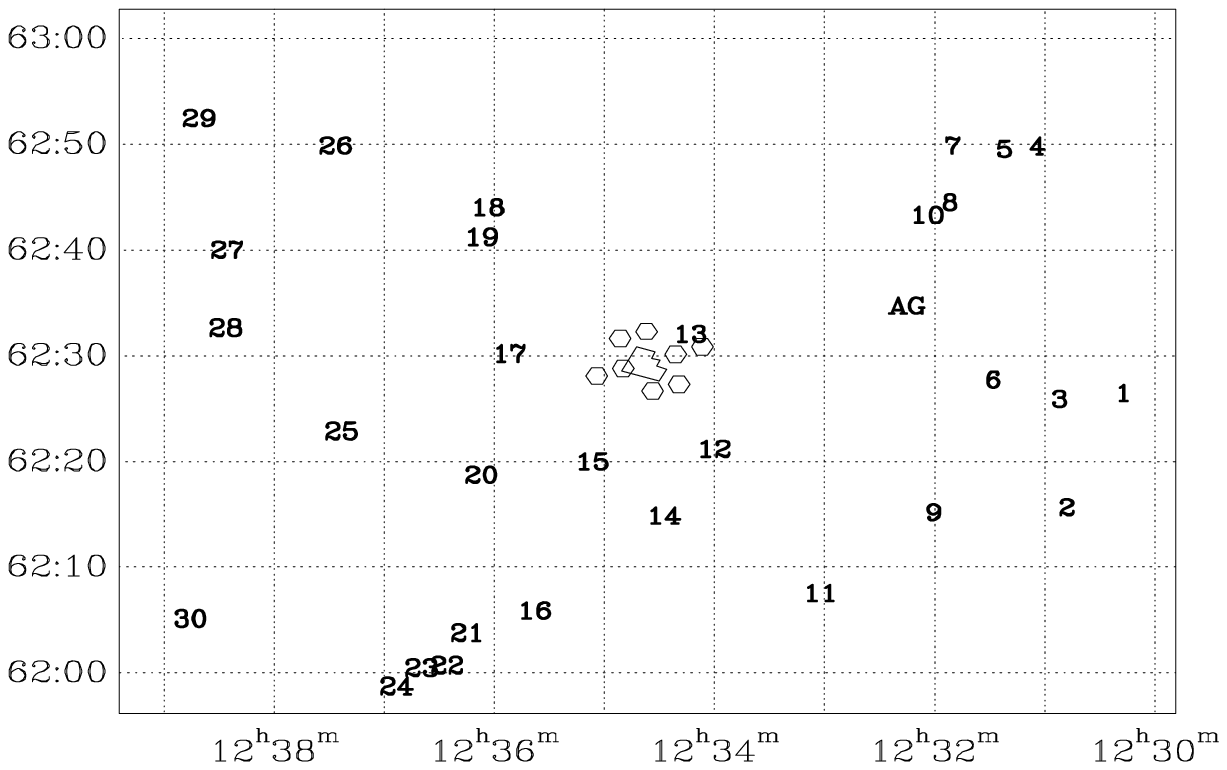}{6in}{0}{100.}{140.}{-324}{-360}
\caption{Schematic map of the QSOs identified in Table 1, relating their
approximate positions with the Hubble Deep Field at the center.
The hexagons mark the centers of the HDF flanking fields.  The active
galaxy at z=0.135 is marked as ``AG.'' R.A. and Dec. are B1950.0.
}
\end{figure}



\begin{thebibliography}{}

\bibitem[]{} Boyle, B.J., Shanks, T., \& Peterson, B.A. 1988, \mnras, 235, 935

\bibitem[]{} Cen, R., Phelps, S., Miralda-Escud\'{e}, J., \& Ostriker, J.P. 
	     1998, \apj, 496, 577

\bibitem[]{} Cohen, J.G., Cowie, L.L., Hogg, D.W., Songaila, A., Blandford, 
	     R.D., Hu, E.M., \& Shopbell, P. 1996, \apj, 471, L5

\bibitem[]{} Cohen, J.G., Blandford, R.D., Hogg, D.W., Pahre, M.A., \& 
             Shopbell, P.L.  1999, \apj, 512, 30

\bibitem[]{} Dinshaw, N., \& Impey, C.D. 1996, \apj, 458, 73

\bibitem[]{} Hall, P.B., Osmer, P.S., Green, R.F., Porter, A.C., \& Warren, 
	     S.J. 1996, \apj, 462, 614

\bibitem[]{} Hernquist, L., Katz, N., Weinberg, D., and Miralda-Escud\'{e}, J. 
	     1996, \apj, 457, 57

\bibitem[]{} Kennefick, J.D., Osmer, P.S., Hall, P.B., \& Green, R.F. 1997, 
             \aj, 114, 2269

\bibitem[]{} Koo, D.C., \& Kron, R.G. 1988, \apj, 325, 92

\bibitem[]{} Koo, D.C., Kron, R.G., \& Cudworth, K.M. 1986, \pasp, 98, 285
 
\bibitem[]{} Livio, M., Fall, S.M., \& Madau, P. 1998, editors, The Hubble 
             Deep Field: Proceedings of the Space Telescope Science Institute 
	     Symposium (Cambridge: Cambridge University Press)

\bibitem[]{} Massey, P., \& Gronwall, C. 1990, \apj, 358, 344

\bibitem[]{} Meylan, G. 1995, editor, QSO Absorption Lines: Proceedings of 
	     the ESO Workshop (Berlin: Springer-Verlag)

\bibitem[]{} Miralda-Escud\'{e}, J., Cen, R., Ostriker, J.P., and Rauch, M. 
	     1996, \apj, 471, 582

\bibitem[]{} Quashnock, J.M., \& Vanden Berk, D.E. 1998, \apj, 500, 28

\bibitem[]{} Rauch, M. 1998, \araa, 38, 267

\bibitem[]{} Sargent, W.L.W., \& Steidel, C.C. 1987, \apj, 322, 142

\bibitem[]{} Valdes, F. 1982, FOCAS User's Manual (Tucson: Kitt Peak National 
	     Observatory)

\bibitem[]{} Warren, S.J., Hewett, P.C., Irwin, M.J., \& Osmer, P.S. 1991, 
	     \apjs, 76, 1

\bibitem[]{} Williams, R.E. et al. 1996, \aj, 112, 1335.

\bibitem[]{} Williams, R.E. et al. 1998, Bulletin of the AAS, 193, 7501

\bibitem[]{} Williger, G.M., Hazard, C., Baldwin, J.A., \& McMahon, R.G. 1996,
	     \apjs, 104, 145

\end{thebibliography}
\end{document}